\documentclass[final,3p,times,twocolumn]{elsarticle}

\usepackage{amssymb}
\usepackage{amsmath}
\usepackage {longtable}
\usepackage{bm}
\usepackage{slashed}

\newcount\hour\newcount\minute
        \hour=\time \divide\hour by60 \minute=\time
        {\multiply\hour by60 \global\advance\minute by-\hour}
        \edef\militarytime{\number\hour:\ifnum\minute<10
0\fi\number\minute}

\newcommand\Refe[1]     {Ref.\,\cite{#1}}

\newcommand\fig[1]     {Fig.\,{\ref{#1}}}
\newcommand\figs[2]    {Figs.\,{\ref{#1}} and ~\ref{#2}}

\newcommand\tab[1]     {Table~\ref{#1}}
\newcommand\tabs[2]    {Tables~{\ref{#1}} and ~\ref{#2}}

\newcommand{\beq}{\begin{equation}}
\newcommand{\eeq}{\end{equation}}
\newcommand{\bea}{\begin{eqnarray}}
\newcommand{\eea}{\end{eqnarray}}
\newcommand{\bc}{\begin{center}}
\newcommand{\ec}{\end{center}}
\def\bsp#1\esp{\begin{split}#1\end{split}}
\def\bal#1\eal{\begin{align}#1\end{align}}
\newcommand{\helac}{\texttt{HELAC}}
\newcommand{\helacdipoles}{\texttt{HELAC-Dipoles}}
\newcommand{\helaconeloop}{\texttt{HELAC-Oneloop}}
\newcommand{\mcatnlo}{\texttt{MC@NLO}}
\newcommand{\pythia}{\texttt{PYTHIA}}
\newcommand{\herwig}{\texttt{HERWIG}}
\newcommand{\madgraph}{\texttt{MADGRAPH}}
\newcommand{\powheghelac}{\texttt{POWHEG+HELAC}}
\newcommand{\powhel}{\texttt{PowHel}}
\newcommand{\powhegbox}{\texttt{POWHEG BOX}}
\newcommand{\pb}{\,\mathrm{pb}}
\newcommand{\fb}{\,\mathrm{fb}}

\newcommand{\gev}{\ensuremath{\,\mathrm{GeV}}}
\newcommand{\tev}{\ensuremath{\,\mathrm{TeV}}}

\newcommand{\ttjet}{t\,$\bar{{\rm t}}$ + jet }
\newcommand{\pT}{\ensuremath{p_{\perp}}}
\newcommand{\lp}{\ensuremath{\ell^+}}
\newcommand{\lm}{\ensuremath{\ell^-}}
\newcommand{\pTlp}{\ensuremath{p_{\perp}^{\ell^+}}}
\newcommand{\pTlm}{\ensuremath{p_{\perp}^{\ell^-}}}
\newcommand{\pTj}{\ensuremath{p_{\perp}^{j}}}

\newcommand{\pTt}{\ensuremath{p_{{\rm t}\perp}}}
\newcommand{\pTtb}{\ensuremath{p_{\bar{\rm t}\perp}}}
\newcommand{\ETmiss}{\ensuremath{E_{\perp}^{\rm miss}}}
\newcommand{\HT}{\ensuremath{H_{\perp}}}
\newcommand{\kT}{\ensuremath{k_{\perp}}}
\newcommand{\mT}{\ensuremath{m_{\perp}}}
\newcommand{\mt}{\ensuremath{m_{\rm t}}}

\newcommand{\muR}{\ensuremath{\mu_{\rm R}}}
\newcommand{\muF}{\ensuremath{\mu_{\rm F}}}

\newcommand{\rt}{{\rm t}}
\newcommand{\bt}{\ensuremath{\bar{{\rm t}}}}
\newcommand{\bq}{\ensuremath{\bar{q}}}
\newcommand{\ptc}{\ensuremath{p_{\perp}^{\rm t.c.}}}
\newcommand{\LO}{{\rm LO}}
\newcommand{\NLO}{{\rm NLO}}

\journal{Elsevier}

\begin{document}
\begin{frontmatter}

\title{Top quark pair production in association with a jet at NLO
accuracy with parton showering}
\author[NCRS,DE]{Adam Kardos}
\author[NCRS]{Costas G.~Papadopoulos}
\author[DE,ATOMKI]{Zolt\'an Tr\'ocs\'anyi}
\address[NCRS]{NCSR Demokritos, Institute of Nuclear Physics, Athens, Greece}
\address[DE]{Institute of Physics, University of Debrecen,\\
H-4010 Debrecen P.O.Box 105, Hungary}
\address[ATOMKI]{Institute of Nuclear Research of the Hungarian
Academy of Sciences, Hungary}
\date{\today}

\begin{abstract}
We compute the production cross section of a top-antitop pair in
association with a jet at hadron colliders at next-to-leading order
accuracy matched with parton shower algorithms to make predictions at
the hadron level. The parton shower allows for including the decay of
the top quarks at the leading order accuracy. We use a framework based
on three well established numerical codes, the \powhegbox,
used for the calculation of the cross section, \helac, which
generates the matrix elements for the Born-level, real emission and the
virtual part, and finally a parton shower program, such as \pythia\ or
\herwig, which generate the parton-shower and hadronization.

\noindent {\em PACS:} 12.38.-t, 13.87.-a, 14.65.Ha
\end{abstract}

\begin{keyword}
QCD \sep jets \sep top quarks
\end{keyword}

\end{frontmatter}

\section{Introduction}

With the startup of the LHC, high energy particle physics entered a new
era. At higher energies, measurements with higher precision become
available, which poses new demands to the theoretical predictions: the
corresponding cross sections are needed beyond leading order (LO)
accuracy even for large multiplicity final states. By now standard
techniques exist \cite{Berger:2008sj,vanHameren:2009dr} for computing
the next-to-leading order (NLO) corrections to many phenomenologically
interesting processes involving four, or more hard objects (heavy
particle or hard jet) in the final state
\cite{Berger:2009ep,Bevilacqua:2009zn,Bevilacqua:2010ve,Berger:2010zx}.
Despite the improved accuracy obtained by computing the cross
sections at NLO, there is still a large gap between fixed order
theoretical predictions and data collected by the detectors. At fixed
order we calculate only hard parton-level processes, while in
experiments we observe hadrons.  The common practice to fill this gap is
the use of parton shower programs \cite{Sjostrand:1993yb,Corcella:2000bw}
which also include hadronization models. The advantage of these
programs is the generation of unweighted events, which can be utilized
for performing the same analysis as on the collected data, allowing for
a direct comparison of theory and experiment, or predicting the
Standard Model background. However, these programs catch only the
important features of small angle radiation off partons, and the
distributions of observable quantities are not expected to give a good
description in the regions dominated by large-angle hard emissions.

Until recent years, these two main approaches were used separately
for making predictions. Merging NLO computations with parton showers
was pioneered by the MCatNLO project \cite{Frixione:2002ik}. By now
all interesting $2 \to 2$ processes are included in the
\mcatnlo\ code \cite{Frixione:2010wd}.  Another method for
merging NLO computations with parton showers, which produces only
positive weight events, was developed in
refs.~\cite{Nason:2004rx,Frixione:2007vw}.  The latter procedure was
later implemented in the \powhegbox\ \cite{Alioli:2010xd}%
\footnote{http://powhegbox.mib.infn.it}. The \powhegbox\ can almost be
considered a black box that requires matrix elements as input and
produces unweighted events in the form of Les-Houches accord files
\cite{Boos:2001cv} as output. These events can be processed with the
\powhegbox\ for generating the showered events for further analysis.

In this letter we show the first application to a $2 \to 3$ process of
the combination of the \powhegbox\ and the \helac\
\cite{vanHameren:2009dr} frameworks for producing showered events of the
\ttjet\ final state that can be used to make distributions with correct
perturbative expansion up to NLO accuracy. Due to the large collision
energy at the LHC, t$\bar{\rm t}$ pairs with large transverse momentum 
will be copiously produced and the probability for the top quarks to
radiate gluons will be sufficiently large to make the \ttjet\ final state
measurable with high statistics. Therefore, we make first predictions for
such events at the TeVatron and the LHC. A more detailed analysis will
be presented elsewhere.

\vspace*{-7pt}
\section{Method}

The cornerstone of our program is the \powhegbox\ \cite{Alioli:2010xd}
framework, that uses the FKS subtraction scheme \cite{Frixione:1995ms}
for the NLO calculation. The \powhegbox\ requires the following input:
\begin{itemize}
\itemsep -2pt
\item
We use flavour structures given in \tabs{tab:Bflavours}{tab:Rflavours}.
\begin{table}
\begin{center}\begin{tabular}{|l|l|l|l|}
\hline
\hline
$qg\to\rt\bt q$ & $gq\to\rt\bt q$ & $\bq g\to\rt\bt \bq $ & $g\bq\to\rt\bt\bq$\\
$gg\to \rt\bt g$ & $q\bq\to \rt\bt g$ & $\bq q\to \rt\bt g$  & \\
\hline
\hline
\end{tabular}
\end{center}
\caption{Flavour structures of the Born processes, $q =$ u, d, c, s, b.}
\label{tab:Bflavours}
\end{table}
\begin{table}
\begin{center}
\begin{tabular}{|l|l|l|}
\hline
\hline
$qg\to\rt\bt qg$ & $qq\to\rt\bt qq$ & $q\bq \to\rt\bt q\bq $ \\
$gq\to\rt\bt qg$ & $\bq \bq \to\rt\bt \bq \bq $ & $\bq q\to\rt\bt q\bq $ \\
$\bq g\to\rt\bar{t}\bq g$ & $q\bq \to\rt\bt gg$  & $q\bq \to\rt\bt q'\bq'$ \\
$g\bq \to\rt\bt \bq g$   & $\bq q\to\rt\bt gg$   & $\bq q\to\rt\bt q'\bq'$ \\
$qq'\to\rt\bt qq'$ & $q\bq'\to\rt\bt q\bq'$  & $gg\to\rt\bt gg$ \\
$\bq q'\to\rt\bt \bq q'$ & $\bq \bq'\to\rt\bt \bq \bq'$ & $gg\to\rt\bt q\bq $ \\
\hline
\hline
\end{tabular}
\end{center}
\caption{Flavour structures of the real-emission processes,
$q, q' =$ u, d, c, s, b.}
\label{tab:Rflavours}
\end{table}
\item
We generate a Born phase space of a massless and two massive momenta
using two two-particle invariants and three angles.
\item
We use \helacdipoles\ \cite{Czakon:2009ss} to calculate all the
tree-level helicity amplitudes for the Born subprocesses
$\rt\,\bt \,g\,g\,g \to 0$ and $\rt\,\bt \,q\,{\bar q}\,g \to 0$ and
the real emission subprocesses $\rt\,\bt \,g\,g\,g\,g \to 0$,
$\rt\,\bt \,q\,{\bar q}\,g\,g \to 0$ and 
$\rt\,\bt \,q\,{\bar q}\,q'\,{\bar q'} \to 0$.  (We define the
corresponding crossing symmetric amplitudes for all incoming momenta
and cross into the relevant physical channels.)
\item
For the colour-correlated squared matrix elements of the Born
subprocesses we use \helacdipoles.
\item
We use the polarization vectors to project the helicity amplitudes to
Lorentz basis for writing the spin-correlated squared matrix elements.
\item
Finally, we obtain the one-loop corrections to the Born subprocesses
utilizing the \helaconeloop\ implementation
\cite{vanHameren:2009dr,Ossola:2007ax,Bevilacqua:2010mx}
of unitary-based numerical evaluation of one-loop amplitudes
\cite{Bern:1994cg,Brandhuber:2005jw,Anastasiou:2006gt,Ossola:2006us,
Ellis:2007br,Bern:2007dw,Ossola:2008xq,Draggiotis:2009yb}.
\end{itemize}
With this input \powhegbox\ can be used to generate hadronic events. One
may choose any parton shower (PS) Monte Carlo program for generating parton
showers, decays of heavy quarks and hadronization. There is one important
point in choosing the PS. We generate events with hardest emission
measured by the transverse momentum of the emission. 
If the ordering variable in the shower is different from the transverse
momentum of the parton splitting (for instance, the angular ordered
showers in \herwig), then the hardest emission is not necessarily the
first one. In such cases the \herwig\ discards shower evolutions (vetoed
shower) with larger transverse momentum in a subsequent splitting than
that in the real emission correction. In addition, a truncated shower
simulating wide-angle soft emission before the first emission is also
needed in principle, but its effect was found small
\cite{LatundeDada:2006gx}. As there is no implementation of truncated
shower in \herwig\ using external LHE event files, the effect of the
truncated showers is absent from our predictions.

\vspace*{-7pt}
\section{Checks}

In order to ensure the correctness of the calculations we performed the
following checks relevant to any fixed order calculation at the NLO
accuracy:
(i) Compared the cross section at LO to the prediction of the public code
\madgraph\ \cite{Alwall:2007st} and found complete agreement.
(ii) 
Checked the virtual correction obtained from the \helaconeloop\ program
in several randomly chosen phase space points to that obtained from
the implementation in the \powhel\ (= \powheghelac) program.
(iii) 
Checked in several randomly chosen phase-space regions that the ratio
of the soft- and collinear limits of the real-emission matrix elements
and subtractions tend to one in all possible unresolved limits.

There is an important technical issue related to the way of calculation
organized in the \powhegbox. The selection cuts are applied on the events
obtained after hadronization. However, when computing the \ttjet\
production cross section at fixed order, the cuts are applied at the
parton level. At LO this means a cut on the transverse momentum of the
only massless parton in the final state. At NLO the virtual contribution
has the same event configuration as the Born one, but the real emission
contribution has two massless partons in the final state, that have to be
combined into a jet before the physical cut can be applied. In the
\powhegbox\ such a separation of the real and virtual contributions is
not possible because the event-generation starts with an underlying Born
configuration from which further parton emissions are generated. In order
to make the parton-level calculation finite, we can apply a technical
cut on the transverse momentum of the single massless parton in the Born
confiuration. With a given set of selection cuts, one has to check that
the chosen technical cut is sufficiently loose such that it does not
influence the physical cross section. Typically we find that for jet
transverse momentum cuts of several tens of GeV, a several GeV technical
cut on the transverse momentum of the massless parton at Born level is
sufficiently loose.  Another way of treating the same problem, also
implemented in the \powhegbox, is to use a suppression factor on the
underlying Born configuration \cite{Alioli:2010qp}.

\vspace*{-7pt}
\section{Comparison to predictions at NLO}

The first calculation of the \ttjet\ production cross section was
computed by Dittmaier, Uwer and Weinzierl
\cite{Dittmaier:2007wz,Dittmaier:2008uj}. In order to further check our
program, we computed the production cross section at NLO accuracy using
the same physical parameters as in \Refe{Dittmaier:2008uj}. Due to the
technical cut mentioned in the previous section, the \powhel\
framework is not optimal for a fixed-order computation, nevertheless
our prediction, $\sigma^{\NLO} = (1.78 \pm 0.01)\pb$
is in agreement with the cross section quoted in \Refe{Dittmaier:2008uj},
$\sigma^{\NLO}= (1.791 \pm 0.001)\pb$, within the uncertainty of our
integration.  Our prediction is independent of the technical cut below
$\ptc \lesssim 5\gev$ as shown in \tab{tab:cutdep}.
\begin{table}
\begin{center}
\begin{tabular}{|c|c|c|}
\hline
\hline
\ptc [GeV] & $\sigma^{\LO}$ [pb]  & $\sigma^{\NLO}$ [pb] \\
\hline
\hline
20 & 1.583 & $1.773 \pm 0.003$ \\
\hline
5 & 1.583 & $1.780 \pm 0.006$ \\
\hline
1 & 1.583 & $1.780 \pm 0.010$ \\
\hline
\hline
\end{tabular}
\end{center}
\caption{Dependence of the NLO cross section on the technical cut \ptc.}
\label{tab:cutdep}
\end{table}

\begin{figure}[t]
\centerline{\includegraphics[width=1.15\linewidth]{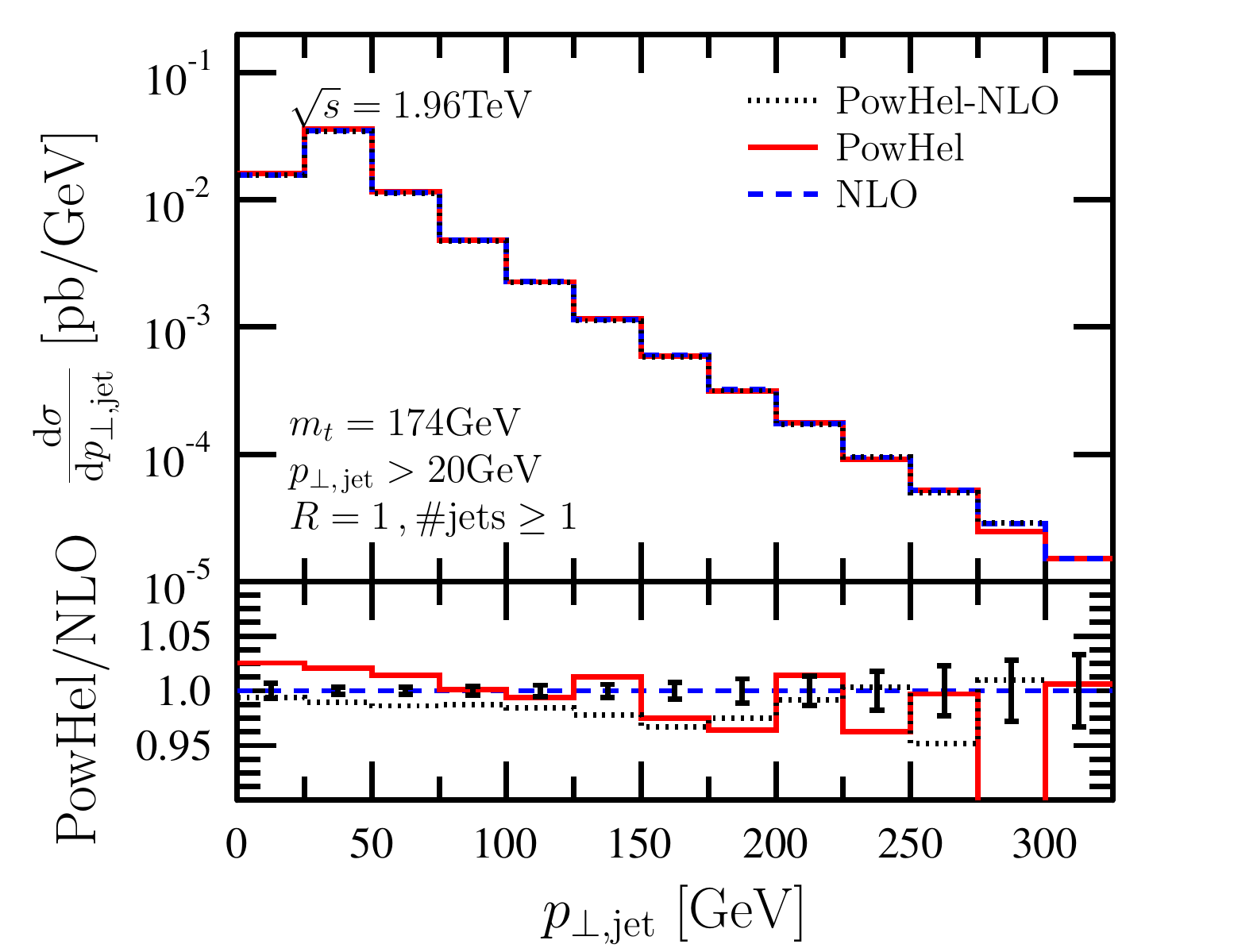}}
\centerline{\includegraphics[width=1.15\linewidth]{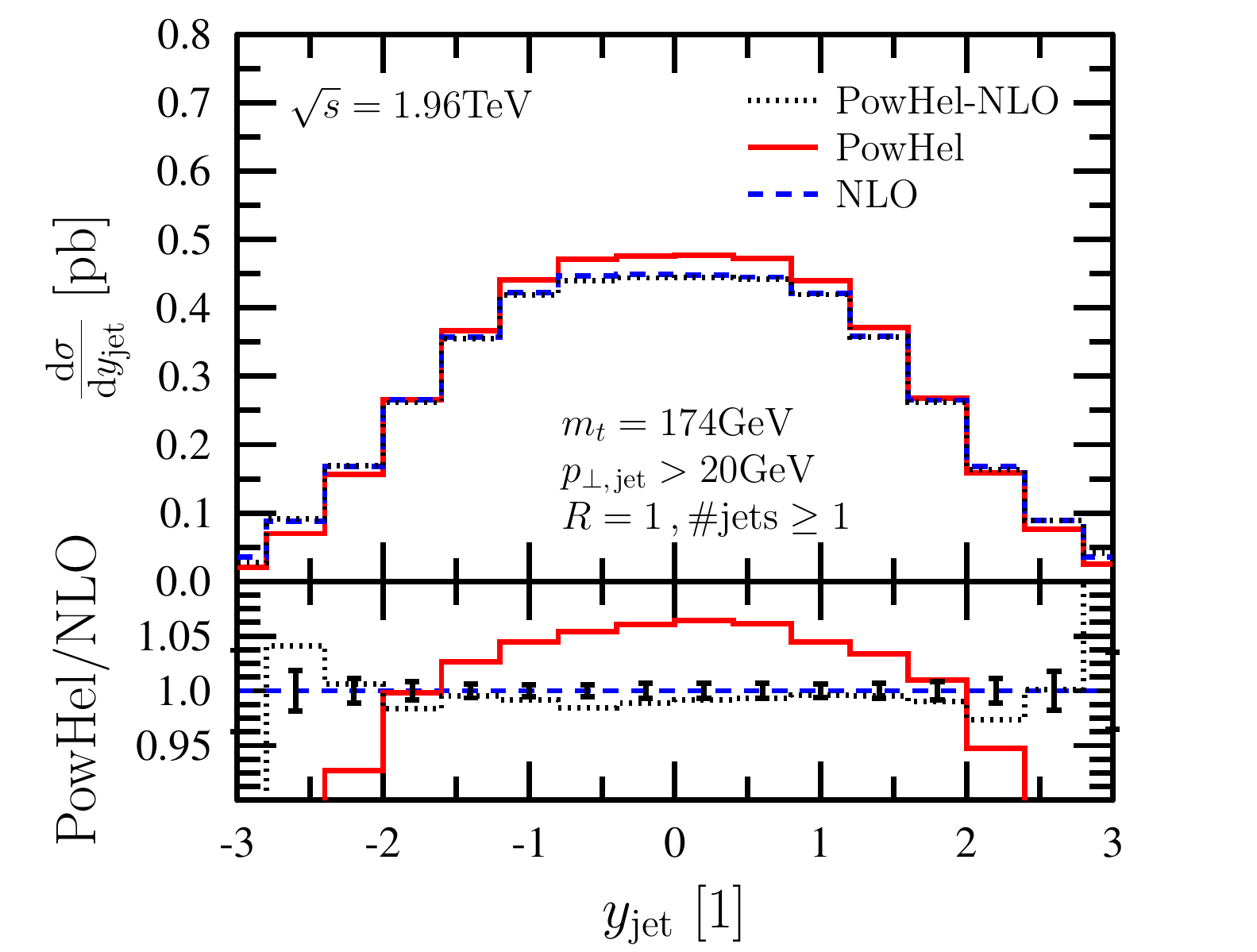}}
\caption{
\label{fig:NLO}
Transverse momentum and rapidity distributions of the jet.}
\vspace*{-7pt}
\end{figure}

In order to check the predictions obtained with Born-suppression, we
computed the distributions published in \Refe{Dittmaier:2008uj} at NLO
accuracy and we found agreement. 
Examples are shown in \fig{fig:NLO} for the case of the
transverse momentum and rapidity distributions of the jet. The lower
panels show the ratio of the PowHel-NLO predictions to the predictions
of \Refe{Dittmaier:2008uj}.  The error bars in the lower panel
represent the combined statistical uncertainty of the two computations.

We also compared distributions obtained from LHE events, including the
first radiation only, to predictions at NLO. For the distributions of the
transverse momenta of the jet (see also in \fig{fig:NLO}) and the
top as well as for the rapidity distribution of the top we found
agreement. The rapidity distribution of the jet is more central
from the LHE events than from NLO.

\vspace*{-7pt}
\section{Effects of decays and shower}

The production of \ttjet\ final state at the NLO accuracy together with
decay of the heavy quarks in the narrow-width approximation (at LO
accuracy) has been published recently by Melnikov and Schulze in
\cite{Melnikov:2010iu}.  In our NLO+PS computation decays of heavy
quarks are implemented in the PS, therefore, spin correlations are not
included. In contrast, the narrow-width approximation allows for taking
into account the spin correlations. Thus, in order to see the effect of the
parton shower, we first generated distributions without the shower, but
with decays (we just included on-shell decays of t-quarks, and further
decays of their decay products, if unstable,
turning off any shower and hadronization effect, marked as `Decay'),
then with the full shower Monte Carlo (marked with the name of the SMC).
We compared the total cross section as well as several distributions to
those predictions made for collisions at the Tevatron, $\sqrt{s}=1.96\tev$,
valid at the NLO accuracy. We generated two million events with
\powhel, which were showered with \pythia-6.4.25
\cite{Sjostrand:2006za} and \herwig-6.5.20 \cite{Corcella:2002jc}
subsequently.  For the comparison, we used the semileptonic decay
channel and the following parameters and selection cuts from
\Refe{Melnikov:2010iu}:
(i) mass of the top quark $\mt = 172\gev$; all other Standard Model
parameters as implemented in the PS programs,
(ii) CTEQ6M parton distribution functions,
(iii) \kT-clustering algorithm with $R = 0.5$ and four-momentum
recombination scheme \cite{Catani:1993hr},
(iv) $\muR=\muF=\mt$,
(v) $\pTlp>20\gev$,
(vi) $\ETmiss>20\gev$,
(vii) $\pTj>20\gev$,
(viii) $|y_j| < 2$,
(ix) minimum five jets,
and (x) $\HT > 220\gev$, where \HT\ is the scalar sum of transverse
momenta in the event,
$
\HT = \pTlp + \ETmiss + \sum_j \pTj\,.
$ 
In addition, if the final state after these selection cuts contained one
or more charged
leptons, we rejected the event if the transverse momentum of this lepton
was above 20\gev. This latter requirement is not needed in a fixed order
calculation, but necessary in ours to select the semileptonic channel.
The technical cut was chosen to \ptc = 5\gev.  

The predicted SMC cross sections are very sensitive to the details
of the analysis. We kept the leptons and neutral pions stable, while all
other particles were allowed to be stable or to decay according to the
default implementation in each SMC. Quark masses, as well as
$W$, $Z$ masses and total decay widths, were tuned to the same values
in \pythia\ and \herwig.  On the other hand, each of the two codes was
allowed to compute autonomously partial branching fractions in
different decay channels for all unstable particles and hadrons.
Multiparticle interaction effects were neglected (default in \herwig).
Additionally, the intrinsic $p_T$ spreading of valence partons in
incoming hadrons in \herwig\ was assumed to be 2.5 GeV.

\begin{figure}[t]
\centerline{\includegraphics[width=1.15\linewidth]{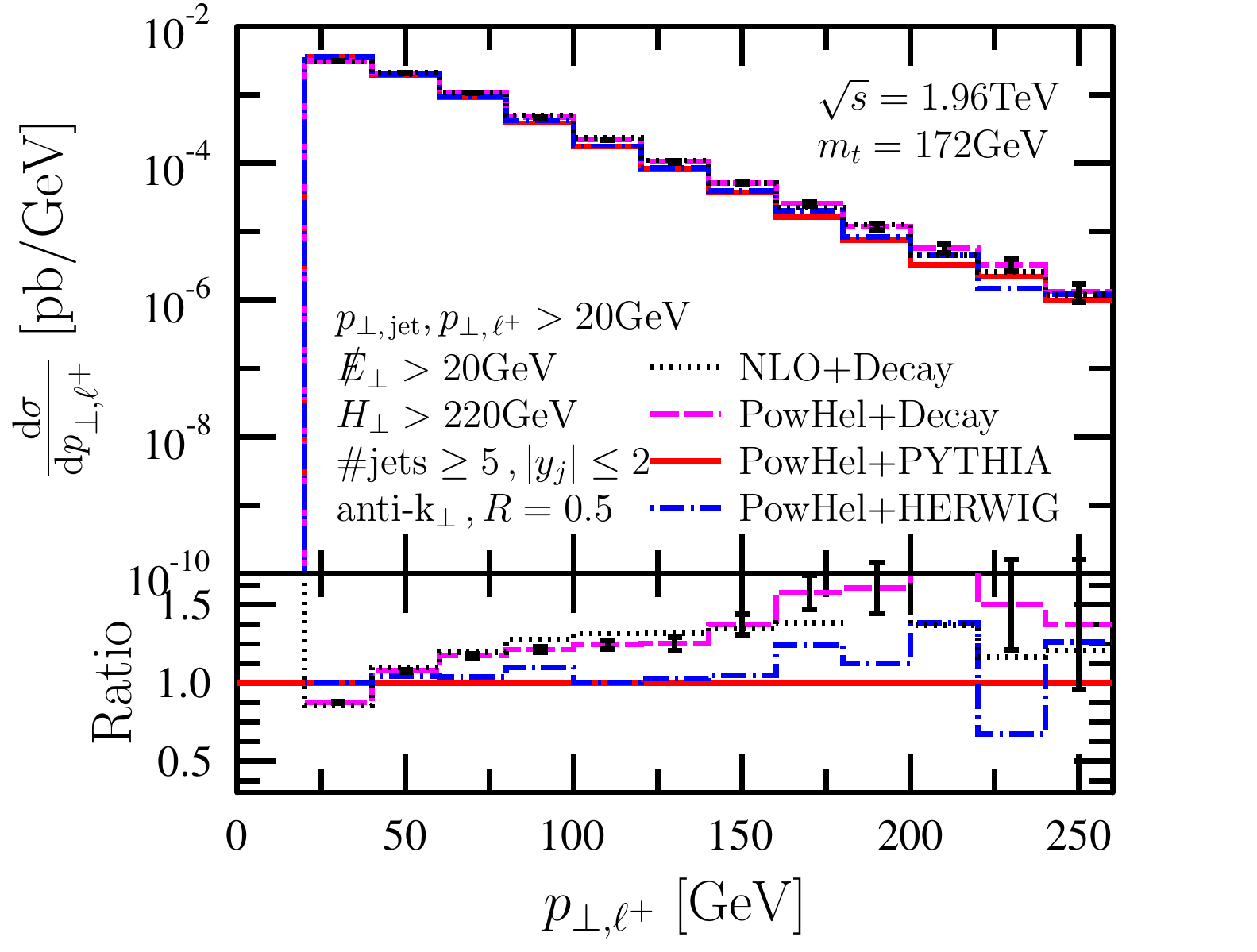}}
\centerline{\includegraphics[width=1.15\linewidth]{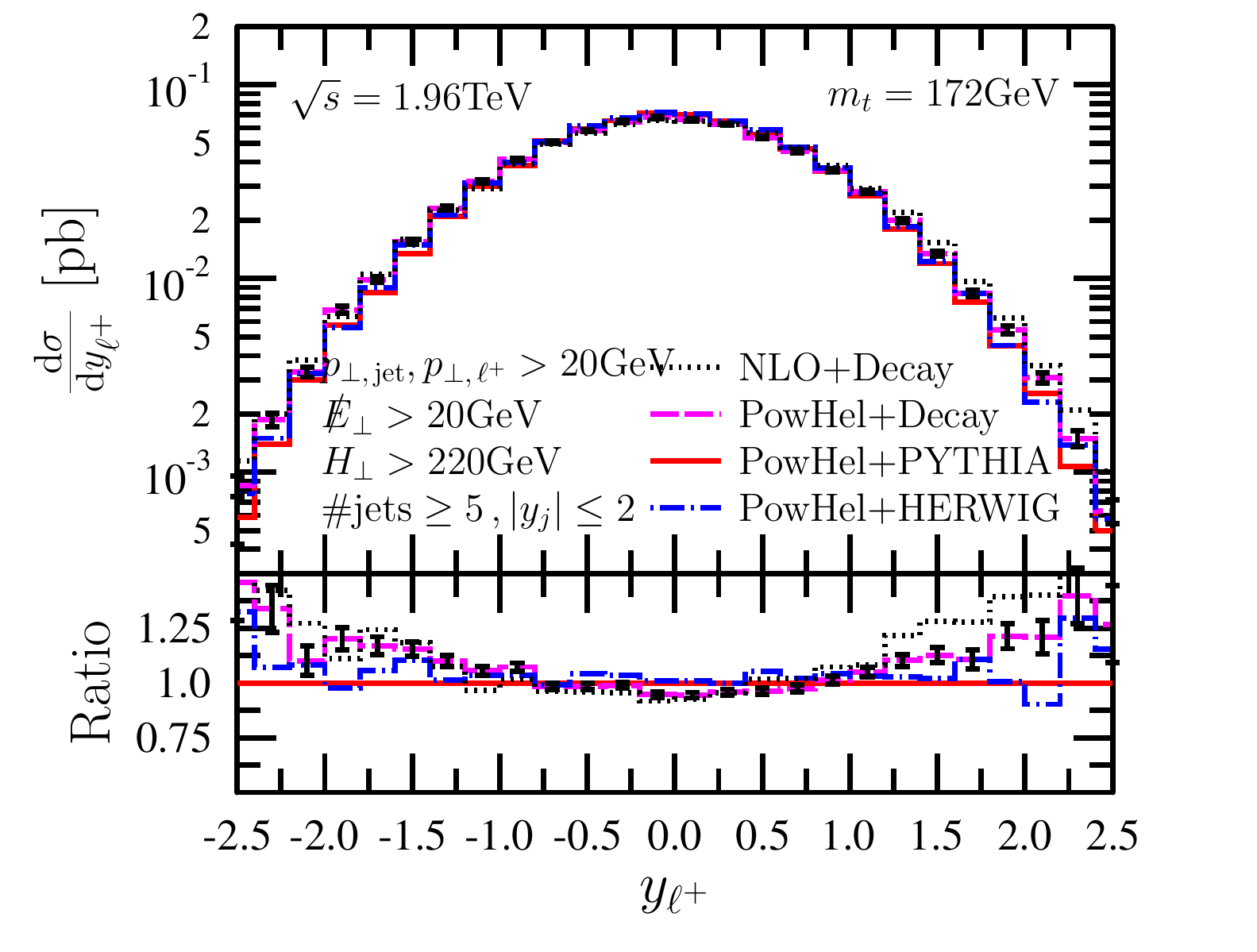}}
\caption{
\label{fig:ylep}
Transverse momentum and rapidity distributions of the antilepton.}
\vspace*{-7pt}
\end{figure}

Considering this setup, we always found agreement between \pythia\ and
\herwig\ predictions within 3\,\%, which is also the effect of
including versus neglecting negative weight events in the analysis. 
For instance, using our selection cuts and taking into
account the negative weight events, we obtained the cross sections
$\sigma^{\powhel+\herwig}=146.9\fb$ and 
$\sigma^{\powhel+\pythia}=143.2\fb$, while
without the negative weight events, we obtain
$\sigma^{\powhel+\pythia}=147\fb$.
The corresponding value for the \powhel+decay case is
$\sigma^{\rm PowHel+decay}=144.2\fb$ (with negative weight events
included).  These numbers cannot be compared
directly to the fixed-order prediction $\sigma^{\rm NLO}= 33.6\fb$
quoted in \Refe{Melnikov:2010iu} for two reasons. On the one hand 
in \Refe{Melnikov:2010iu} only one lepton family was considered
in the decay of the t-quarks, while our prediction contains all three
families. We checked that taking into account only one lepton family in
the decay we obtain a factor of three reduction of the cross section as
expected. On the other hand the authors of \Refe{Melnikov:2010iu} also
observed that there is a large contribution to the cross section from
the emission of a hard jet from the top decay products (estimated an
additional 60\,\% at LO \cite{Schulze:private}), which is included in
our calculation, but not in their value. As this effect is not known at
the NLO accuracy, in order to compare only the shapes of distributions
with only decays included, we multiply the NLO predictions with $r =
\sigma^{\rm PowHel+decay}/\sigma^{\rm NLO}=4.29$
(shown as `NLO+decay' in \figs{fig:ylep}{fig:ht}). The lower panels show the
ratio of the various predictions to the \powhel+\pythia\ one. In order to
exhibit the size of the statistical uncertainty (corresponding to two
million LHE events), avoiding at the same time a very confusing plot, we
show the uncertainty of only the \powhel+decay prediction with errorbars.

In \fig{fig:ylep} we compare the transverse momentum and rapidity
distributions of the antilepton at several different levels.
We observe on these plots some general features:
(i) the two \powhel+SMC predictions are very close except in bins with
low statistics;
(ii) the \powhel+decay predictions are very close to the NLO ones in the
central rapidity region and for the whole $\pT$ range. Looking more
closely, we find that the spin correlations make the NLO rapidity
distribution slightly wider. The addition of the parton shower makes the
rapidity distribution a little even more central due to soft leptons
emitted by the shower in central regions.  (For jet rapidities, not
shown here, the NLO and \powhel+decay predictions coincide, but the shower
effect is much more pronounced.) The $\pT$-distributions of the leptons
becomes much softer for the same reason. The same applies to the
\pT\-spectra of the jets.

\begin{figure}[t]
\centerline{\includegraphics[width=1.15\linewidth]{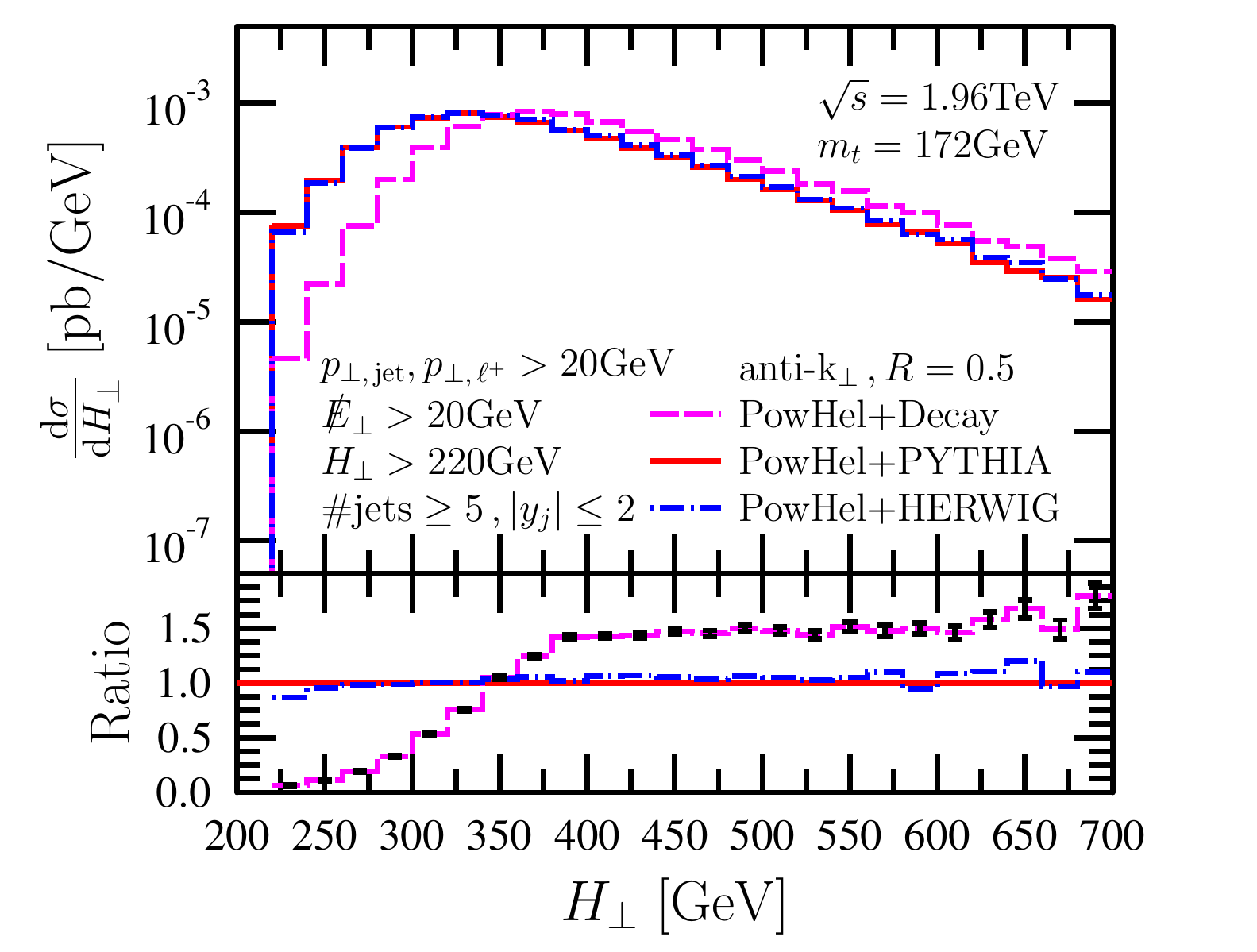}}
\caption{
\label{fig:ht}
Distribution of the scalar sum of transverse momenta.}
\vspace*{-7pt}
\end{figure}

We find even larger shower effects in the comparison of the
\HT-distributions in \fig{fig:ht} at the decay and SMC levels.  The
shower makes the distribution softer, readily understood as the effect
of unclustered soft hadrons in the event, that appear only in the shower. 

\vspace*{-7pt}
\section{Predictions for the LHC}

We now turn our attention to the LHC and make some predictions for the
inclusive \ttjet\ production at the low-energy run, $\sqrt{s} = 7\tev$
in the dileptonic final state channel. We apply the following selection
criteria:
(i) at least three jets 
are reconstructed with the anti-\kT-clustering algorithm with $R = 0.5$
and four-momentum recombination scheme \cite{Cacciari:2005hq},
(ii) $\pTj>30\gev$,
(iii) $|y_j| < 2.5$,
(iv) $\ETmiss>30\gev$ for $e^+e^-$ and $\mu^+\mu^-$ pairs, while
$\ETmiss>20\gev$ for $e^\pm\mu^\mp$ pairs,
(v) $\pTlm$, $\pTlp>20\gev$
for exactly one \lp\ and one \lm.

For default scales we used two different choices: (i) the mass of the
t-quark, \mt, and (ii) the transverse mass of the harder top,
$\muR=\muF=\mT$, where
$
\mT=\sqrt{\mt^2+\max\{\pTt^2,\pTtb^2\}}
$. 
We expect the latter scale better interpolates between near-threshold
and hard events.

\begin{figure}[t]
\centerline{ \includegraphics[width=1.15\linewidth]{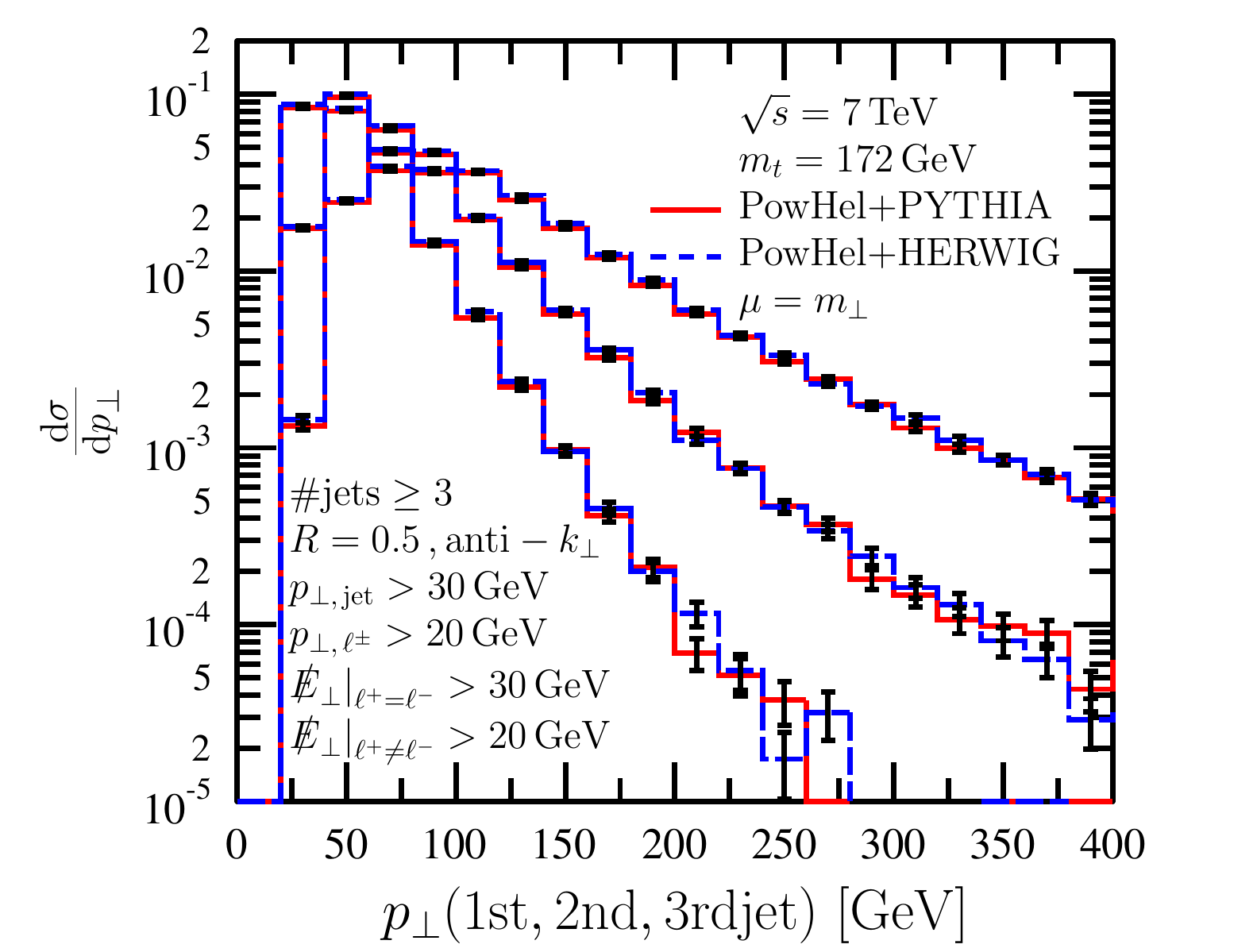}}
\caption{\label{fig:pTjets} 
Transverse momentum distributions of the first, second and third hardest
jet.}
\vspace*{-7pt}
\end{figure}

\begin{figure}[t]
\centerline{ \includegraphics[width=1.15\linewidth]{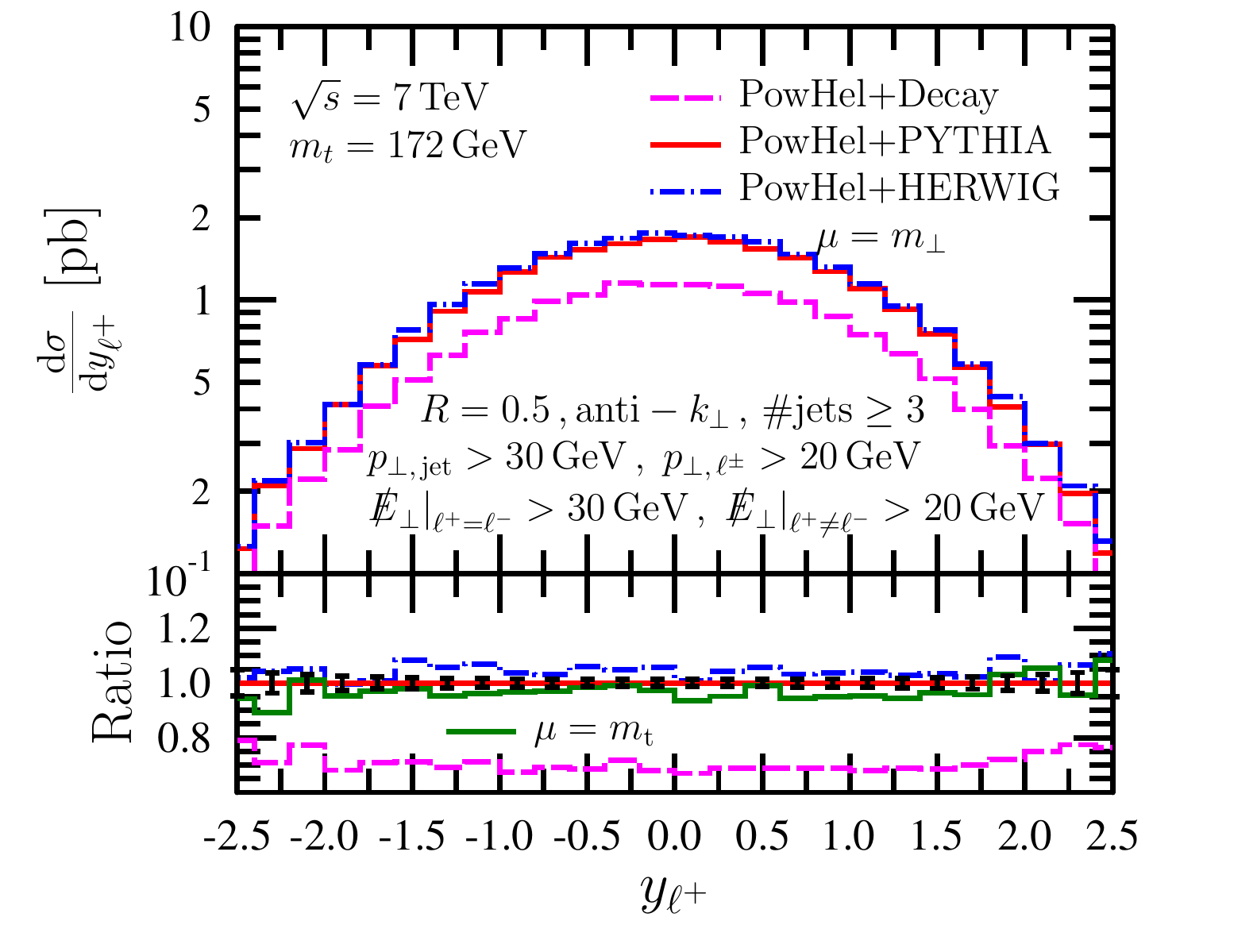}}
\caption{\label{fig:yleplhc} Rapdity distribution of the antilepton.
The lower plot also includes the ratio of the cross section obtained
with $\mu = \mt$ to that obtained with $\mu = \mT$ (\powhel+\pythia).}
\vspace*{-7pt}
\end{figure}

\begin{figure}[t]
\centerline{ \includegraphics[width=1.15\linewidth]{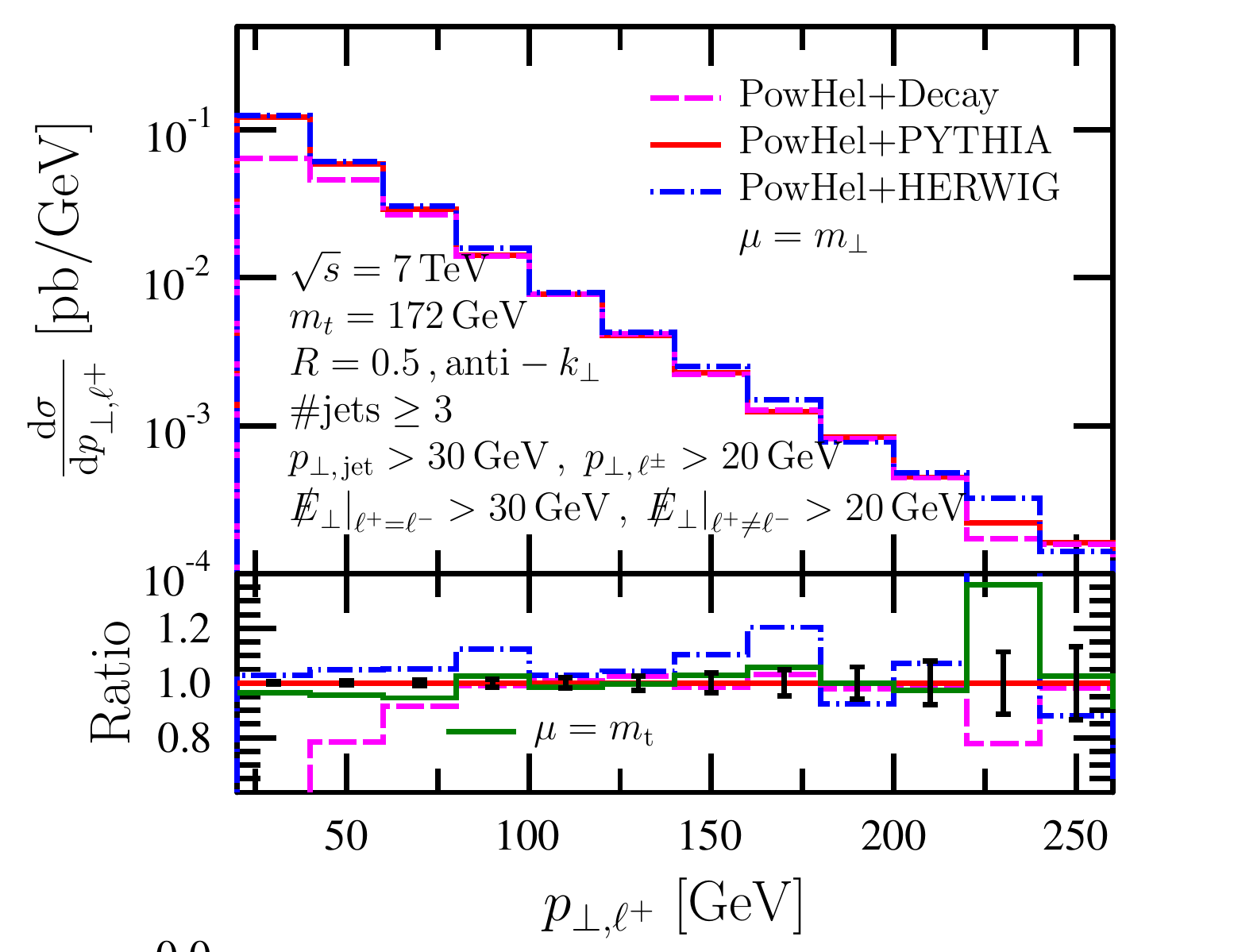}}
\caption{\label{fig:pTlep} 
Transverse momentum distribution of the antilepton. The lower plot also
includes
the ratio of the cross section obtained with $\mu = \mt$ to that
obtained with $\mu = \mT$ (\powhel+\pythia).}
\vspace*{-7pt}
\end{figure}

In \fig{fig:pTjets} we plot the transverse momentum distributions of the
hardest, second hardest and third hardest jet.  These \pT\ spectra are
insensitive to the version of the parton shower within the statistical
uncertainty of the computations, which shows that the effect of the
missing truncated shower must be small. Also they are rather robust
against the choice of the default scale (2--6\,\% variation, not shown
here), suggesting small scale dependence in general, but we shall study
that in a separate publication. The same features are also true for the
rapidity distribution of the antilepton as seen in \fig{fig:yleplhc},
where we also exhibited the prediction at the decay level. The lower
panel shows the ratios of the
predictions to the \powhel+\herwig\ case. The error bars represent the
statistical uncertainty of the latter only. We find large (almost
20\,\%) and almost uniform effect of the shower and hadronization.  In
the case of the transverse momentum distribution of the antilepton the
various predictions agree over the whole spectrum except that we see a
large increase from the decay level to the full SMC at small $\pT$, see
\fig{fig:pTlep}. We attribute this increase to the numerous secondary
antileptons generated in the hadronization phases. 


\begin{figure}[t]
\centerline{ \includegraphics[width=1.15\linewidth]{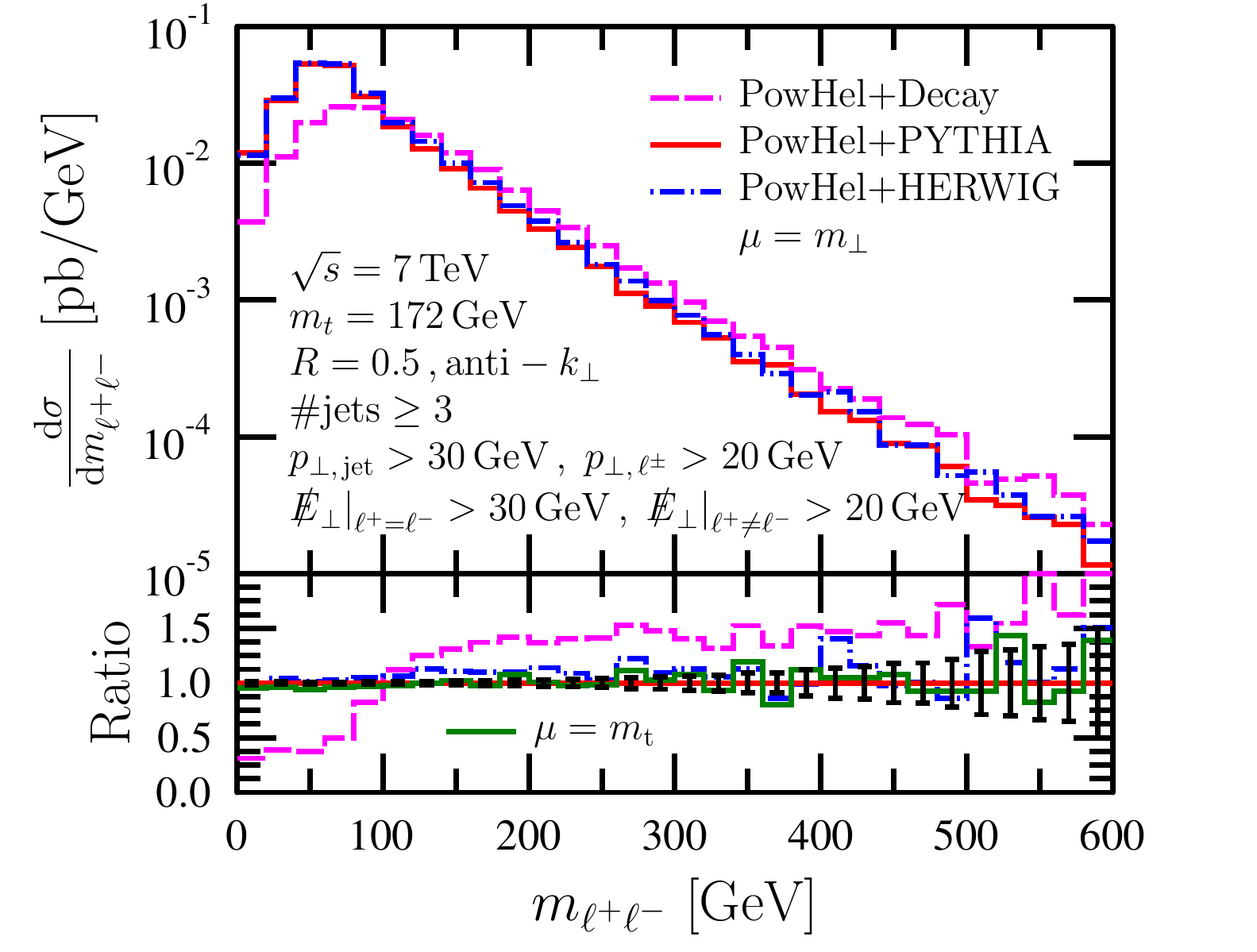}}
\caption{\label{fig:mll} 
Invariant mass distribution of the lepton-antilepton pair. The lower plot
also includes the ratio of the cross section obtained with $\mu =
\mt$ to that obtained with $\mu = \mT$ (\powhel+\pythia).}
\vspace*{-7pt}
\end{figure}

Finally, we plot the invariant mass distribution of the \lp \lm\ pairs
in \fig{fig:mll}.  Here again the full SMC predictions are all the
same. During hadronization additional (anti)leptons with $\pT > 20$\,GeV
may appear and such events are dropped due to our selection cut (v),
resulting in a softer spectrum.

\section{Conclusions}

In this paper we interfaced the \powhegbox\ with the \helac\ framework to 
perform NLO calculations matched with parton showers and hadronization
in a quite general and semi-automatic way. The latter means that the
necessary ingredients for the \powhegbox\ can be taken from the
\helac\ framework without any further computations. We presented
the feasibility with a non-trivial process, namely inclusive \ttjet\
production and decay, and we found reliable results.  We employed
decays as implemented in standard PS Monte Carlo programs.  We leave
the extension to decays included in the hard matrix elements for a
future study.  

We emphasize that the necessary virtual emission was calculated by a
general numerical method which can be used for further processes. Due
to the general nature of our framework including further processes is
feasible.  

Using the \powhel\ framework we produce several million unweighted
events at the hadron level readily available for analysis. These events
can be used to produce distributions that are correct at NLO accuracy
when expanded in the strong coupling. 
Our analyses clearly show the importance of the full SMC. There
are certain regions in the phase space, where even a NLO accuracy is
insufficient. A singular example is the $H_\perp$ distribution which
shows significant softening over the whole kinematic range.

In preparing this letter we learnt about a similar work in progress by
Alioli, Moch and Uwer, presented at the Heavy particles at the LHC
workshop, Zurich, 2011.

This research was supported by
the HEPTOOLS network MRTN-CT-2006-035505,
the LHCPhenoNet network PITN-GA-2010-264564,
the Swiss National Science Foundation Joint Research Project SCOPES IZ73Z0\_1/28079,
and the T\'AMOP 4.2.1./B-09/1/KONV-2010-0007 project.
We are grateful
to P.~Nason and C.~Oleari for their help in using the POWHEG program,
to M.~Schulze for private communications and to M.V.~Garzelli for her
help in revising the analysis.
AK is grateful
to NCSR Demokritos for hospitality and
to G.~Bevilacqua for
useful discussions and help with the original \helac\ programs.
We acknowledge the generous offer of J.~Huston providing us
access to the MSU computer cluster.
\vspace*{-9pt}

\bibliographystyle{elsarticle-num-names}

\begin{thebibliography}{34}
\providecommand{\natexlab}[1]{#1}
\providecommand{\url}[1]{\texttt{#1}}
\providecommand{\urlprefix}{URL }
\expandafter\ifx\csname urlstyle\endcsname\relax
  \providecommand{\doi}[1]{doi:\discretionary{}{}{}#1}\else
  \providecommand{\doi}[1]{doi:\discretionary{}{}{}\begingroup
  \urlstyle{rm}\url{#1}\endgroup}\fi
\providecommand{\bibinfo}[2]{#2}

\bibitem[{Berger et~al.(2008)}]{Berger:2008sj}
\bibinfo{author}{C.~F. Berger}, et~al., \bibinfo{title}{{An Automated
  Implementation of On-Shell Methods for One- Loop Amplitudes}},
  \bibinfo{journal}{Phys. Rev.} \bibinfo{volume}{D78} (\bibinfo{year}{2008})
  \bibinfo{pages}{036003} [arXiv:803.4180].

\bibitem[{van Hameren et~al.(2009)van Hameren, Papadopoulos, and
  Pittau}]{vanHameren:2009dr}
\bibinfo{author}{A.~van Hameren}, \bibinfo{author}{C.~G. Papadopoulos},
  \bibinfo{author}{R.~Pittau}, \bibinfo{title}{{Automated one-loop
  calculations: a proof of concept}}, \bibinfo{journal}{JHEP}
  \bibinfo{volume}{09} (\bibinfo{year}{2009}) \bibinfo{pages}{106}
[arXiv:0903.4665].

\bibitem[{Berger et~al.(2009)}]{Berger:2009ep}
\bibinfo{author}{C.~F. Berger}, et~al., \bibinfo{title}{{Next-to-Leading Order
  QCD Predictions for W+3-Jet Distributions at Hadron Colliders}},
  \bibinfo{journal}{Phys. Rev.} \bibinfo{volume}{D80} (\bibinfo{year}{2009})
  \bibinfo{pages}{074036} [arXiv:0907.1984].

\bibitem[{Bevilacqua et~al.(2009)Bevilacqua, Czakon, Papadopoulos, Pittau, and
  Worek}]{Bevilacqua:2009zn}
\bibinfo{author}{G.~Bevilacqua}, \bibinfo{author}{M.~Czakon},
  \bibinfo{author}{C.~G. Papadopoulos}, \bibinfo{author}{R.~Pittau},
  \bibinfo{author}{M.~Worek}, \bibinfo{title}{{Assault on the NLO Wishlist: pp
  $\to$ tt bb}}, \bibinfo{journal}{JHEP} \bibinfo{volume}{09}
  (\bibinfo{year}{2009}) \bibinfo{pages}{109} [arXiv:0907.4723].

\bibitem[{Bevilacqua et~al.(2010{\natexlab{a}})Bevilacqua, Czakon,
  Papadopoulos, and Worek}]{Bevilacqua:2010ve}
\bibinfo{author}{G.~Bevilacqua}, \bibinfo{author}{M.~Czakon},
  \bibinfo{author}{C.~G. Papadopoulos}, \bibinfo{author}{M.~Worek},
  \bibinfo{title}{{Dominant QCD Backgrounds in Higgs Boson Analyses at the LHC:
  A Study of pp $\to$ t anti-t + 2 jets at Next-To-Leading Order}},
  \bibinfo{journal}{Phys. Rev. Lett.} \bibinfo{volume}{104}
  (\bibinfo{year}{2010}{\natexlab{a}}) \bibinfo{pages}{162002}
[arXiv:1002.4009].

\bibitem[{Berger et~al.(2010)}]{Berger:2010zx}
\bibinfo{author}{C.~F. Berger}, et~al., \bibinfo{title}{{Precise Predictions
  for W + 4 Jet Production at the Large Hadron Collider}}
[arXiv:1009.2338].

\bibitem[{Sjostrand(1994)}]{Sjostrand:1993yb}
\bibinfo{author}{T.~Sjostrand}, \bibinfo{title}{{High-energy physics event
  generation with PYTHIA 5.7 and JETSET 7.4}}, \bibinfo{journal}{Comput. Phys.
  Commun.} \bibinfo{volume}{82} (\bibinfo{year}{1994})
\bibinfo{pages}{74--90}.

\bibitem[{Corcella et~al.(2001)}]{Corcella:2000bw}
\bibinfo{author}{G.~Corcella}, et~al., \bibinfo{title}{{HERWIG 6: an event
  generator for Hadron Emission Reactions With Interfering Gluons (including
  supersymmetric processes)}}, \bibinfo{journal}{JHEP} \bibinfo{volume}{01}
  (\bibinfo{year}{2001}) \bibinfo{pages}{010} [hep-ph/0011363].

\bibitem[{Frixione and Webber(2002)}]{Frixione:2002ik}
\bibinfo{author}{S.~Frixione}, \bibinfo{author}{B.~R. Webber},
  \bibinfo{title}{{Matching NLO QCD computations and parton shower
  simulations}}, \bibinfo{journal}{JHEP} \bibinfo{volume}{06}
  (\bibinfo{year}{2002}) \bibinfo{pages}{029} [hep-ph/0204244].

\bibitem[{Frixione et~al.(2010)Frixione, Stoeckli, Torrielli, Webber, and
  White}]{Frixione:2010wd}
\bibinfo{author}{S.~Frixione}, \bibinfo{author}{F.~Stoeckli},
  \bibinfo{author}{P.~Torrielli}, \bibinfo{author}{B.~R. Webber},
  \bibinfo{author}{C.~D. White}, \bibinfo{title}{{The MCatNLO 4.0 Event
  Generator}}  [arXiv:1010.0819].

\bibitem[{Nason(2004)}]{Nason:2004rx}
\bibinfo{author}{P.~Nason}, \bibinfo{title}{{A new method for combining NLO QCD
  with shower Monte Carlo algorithms}}, \bibinfo{journal}{JHEP}
  \bibinfo{volume}{11} (\bibinfo{year}{2004}) \bibinfo{pages}{040}
[hep-ph/0409146].

\bibitem[{Frixione et~al.(2007)Frixione, Nason, and Oleari}]{Frixione:2007vw}
\bibinfo{author}{S.~Frixione}, \bibinfo{author}{P.~Nason},
  \bibinfo{author}{C.~Oleari}, \bibinfo{title}{{Matching NLO QCD computations
  with Parton Shower simulations: the POWHEG method}}, \bibinfo{journal}{JHEP}
  \bibinfo{volume}{11} (\bibinfo{year}{2007}) \bibinfo{pages}{070}
[arXiv:0709.2092].

\bibitem[{Alioli et~al.(2010)Alioli, Nason, Oleari, and Re}]{Alioli:2010xd}
\bibinfo{author}{S.~Alioli}, \bibinfo{author}{P.~Nason},
  \bibinfo{author}{C.~Oleari}, \bibinfo{author}{E.~Re}, \bibinfo{title}{{A
  general framework for implementing NLO calculations in shower Monte Carlo
  programs: the POWHEG BOX}}, \bibinfo{journal}{JHEP} \bibinfo{volume}{06}
  (\bibinfo{year}{2010}) \bibinfo{pages}{043} [arXiv:1002.2581].

\bibitem[{Boos et~al.(2001)}]{Boos:2001cv}
\bibinfo{author}{E.~Boos}, et~al., \bibinfo{title}{{Generic user process
  interface for event generators}}  [hep-ph/0109068].

\bibitem[{Frixione et~al.(1996)Frixione, Kunszt, and Signer}]{Frixione:1995ms}
\bibinfo{author}{S.~Frixione}, \bibinfo{author}{Z.~Kunszt},
  \bibinfo{author}{A.~Signer}, \bibinfo{title}{{Three jet cross-sections to
  next-to-leading order}}, \bibinfo{journal}{Nucl. Phys.}
  \bibinfo{volume}{B467} (\bibinfo{year}{1996}) \bibinfo{pages}{399--442}
[hep-ph/9512328].

\bibitem[{Czakon et~al.(2009)Czakon, Papadopoulos, and Worek}]{Czakon:2009ss}
\bibinfo{author}{M.~Czakon}, \bibinfo{author}{C.~G. Papadopoulos},
  \bibinfo{author}{M.~Worek}, \bibinfo{title}{{Polarizing the Dipoles}},
  \bibinfo{journal}{JHEP} \bibinfo{volume}{08} (\bibinfo{year}{2009})
  \bibinfo{pages}{085} [arXiv:0905.0883].

\bibitem[{Ossola et~al.(2008{\natexlab{a}})Ossola, Papadopoulos, and
  Pittau}]{Ossola:2007ax}
\bibinfo{author}{G.~Ossola}, \bibinfo{author}{C.~G. Papadopoulos},
  \bibinfo{author}{R.~Pittau}, \bibinfo{title}{{CutTools: a program
  implementing the OPP reduction method to compute one-loop amplitudes}},
  \bibinfo{journal}{JHEP} \bibinfo{volume}{03}
  (\bibinfo{year}{2008}{\natexlab{a}}) \bibinfo{pages}{042}
[arXiv:0711.3596].

\bibitem[{Bevilacqua et~al.(2010{\natexlab{b}})}]{Bevilacqua:2010mx}
\bibinfo{author}{G.~Bevilacqua}, et~al., \bibinfo{title}{{NLO QCD calculations
  with HELAC-NLO}}, \bibinfo{journal}{Nucl. Phys. Proc. Suppl.}
  \bibinfo{volume}{205-206} (\bibinfo{year}{2010}{\natexlab{b}})
  \bibinfo{pages}{211--217} [arXiv:1007.4918].

\bibitem[{Bern et~al.(1995)Bern, Dixon, Dunbar, and Kosower}]{Bern:1994cg}
\bibinfo{author}{Z.~Bern}, \bibinfo{author}{L.~J. Dixon},
  \bibinfo{author}{D.~C. Dunbar}, \bibinfo{author}{D.~A. Kosower},
  \bibinfo{title}{{Fusing gauge theory tree amplitudes into loop amplitudes}},
  \bibinfo{journal}{Nucl. Phys.} \bibinfo{volume}{B435} (\bibinfo{year}{1995})
  \bibinfo{pages}{59--101} [hep-ph/9409265].

\bibitem[{Brandhuber et~al.(2005)Brandhuber, McNamara, Spence, and
  Travaglini}]{Brandhuber:2005jw}
\bibinfo{author}{A.~Brandhuber}, \bibinfo{author}{S.~McNamara},
  \bibinfo{author}{B.~J. Spence}, \bibinfo{author}{G.~Travaglini},
  \bibinfo{title}{{Loop amplitudes in pure Yang-Mills from generalised
  unitarity}}, \bibinfo{journal}{JHEP} \bibinfo{volume}{10}
  (\bibinfo{year}{2005}) \bibinfo{pages}{011} [hep-th/0506068].

\bibitem[{Anastasiou et~al.(2007)Anastasiou, Britto, Feng, Kunszt, and
  Mastrolia}]{Anastasiou:2006gt}
\bibinfo{author}{C.~Anastasiou}, \bibinfo{author}{R.~Britto},
  \bibinfo{author}{B.~Feng}, \bibinfo{author}{Z.~Kunszt},
  \bibinfo{author}{P.~Mastrolia}, \bibinfo{title}{{Unitarity cuts and reduction
  to master integrals in d dimensions for one-loop amplitudes}},
  \bibinfo{journal}{JHEP} \bibinfo{volume}{03} (\bibinfo{year}{2007})
  \bibinfo{pages}{111} [hep-ph/0612277].

\bibitem[{Ossola et~al.(2007)Ossola, Papadopoulos, and Pittau}]{Ossola:2006us}
\bibinfo{author}{G.~Ossola}, \bibinfo{author}{C.~G. Papadopoulos},
  \bibinfo{author}{R.~Pittau}, \bibinfo{title}{{Reducing full one-loop
  amplitudes to scalar integrals at the integrand level}},
  \bibinfo{journal}{Nucl. Phys.} \bibinfo{volume}{B763} (\bibinfo{year}{2007})
  \bibinfo{pages}{147--169} [hep-ph/0609007].

\bibitem[{Ellis et~al.(2008)Ellis, Giele, and Kunszt}]{Ellis:2007br}
\bibinfo{author}{R.~K. Ellis}, \bibinfo{author}{W.~T. Giele},
  \bibinfo{author}{Z.~Kunszt}, \bibinfo{title}{{A Numerical Unitarity Formalism
  for Evaluating One-Loop Amplitudes}}, \bibinfo{journal}{JHEP}
  \bibinfo{volume}{03} (\bibinfo{year}{2008}) \bibinfo{pages}{003}
[arXiv:0708.2398].

\bibitem[{Bern et~al.(2007)Bern, Dixon, and Kosower}]{Bern:2007dw}
\bibinfo{author}{Z.~Bern}, \bibinfo{author}{L.~J. Dixon},
  \bibinfo{author}{D.~A. Kosower}, \bibinfo{title}{{On-Shell Methods in
  Perturbative QCD}}, \bibinfo{journal}{Annals Phys.} \bibinfo{volume}{322}
  (\bibinfo{year}{2007}) \bibinfo{pages}{1587--1634} [arXiv:0704.2798].

\bibitem[{Ossola et~al.(2008{\natexlab{b}})Ossola, Papadopoulos, and
  Pittau}]{Ossola:2008xq}
\bibinfo{author}{G.~Ossola}, \bibinfo{author}{C.~G. Papadopoulos},
  \bibinfo{author}{R.~Pittau}, \bibinfo{title}{{On the Rational Terms of the
  one-loop amplitudes}}, \bibinfo{journal}{JHEP} \bibinfo{volume}{05}
  (\bibinfo{year}{2008}{\natexlab{b}}) \bibinfo{pages}{004}
[arXiv:0802.1876].

\bibitem[{Draggiotis et~al.(2009)Draggiotis, Garzelli, Papadopoulos, and
  Pittau}]{Draggiotis:2009yb}
\bibinfo{author}{P.~Draggiotis}, \bibinfo{author}{M.~V. Garzelli},
  \bibinfo{author}{C.~G. Papadopoulos}, \bibinfo{author}{R.~Pittau},
  \bibinfo{title}{{Feynman Rules for the Rational Part of the QCD 1-loop
  amplitudes}}, \bibinfo{journal}{JHEP} \bibinfo{volume}{04}
  (\bibinfo{year}{2009}) \bibinfo{pages}{072} [arXiv:0903.0356].

\bibitem{LatundeDada:2006gx}
\bibinfo{author}{O.~Latunde-Dada},\bibinfo{author}{S.~Gieseke},
\bibinfo{author}{B.~Webber},
  JHEP {\bf 0702}, 051 (2007)
  [arXiv:hep-ph/0612281].

\bibitem{Alioli:2010qp}
\bibinfo{author}{S.~Alioli}, \bibinfo{author}{P.~Nason},
  \bibinfo{author}{C.~Oleari}, \bibinfo{author}{E.~Re},
  \bibinfo{title}{{Vector boson plus one jet production in POWHEG}},
  \bibinfo{journal}{JHEP} \bibinfo{volume}{01} (\bibinfo{year}{2011})
  \bibinfo{pages}{095} [arXiv:1009.5594].

\bibitem[{Alwall et~al.(2007)}]{Alwall:2007st}
\bibinfo{author}{J.~Alwall}, et~al., \bibinfo{title}{{MadGraph/MadEvent v4: The
  New Web Generation}}, \bibinfo{journal}{JHEP} \bibinfo{volume}{09}
  (\bibinfo{year}{2007}) \bibinfo{pages}{028} [arXiv:0706.2334].

\bibitem[{Dittmaier et~al.(2007)Dittmaier, Uwer, and
  Weinzierl}]{Dittmaier:2007wz}
\bibinfo{author}{S.~Dittmaier}, \bibinfo{author}{P.~Uwer},
  \bibinfo{author}{S.~Weinzierl}, \bibinfo{title}{{NLO QCD corrections to t
  anti-t + jet production at hadron colliders}}, \bibinfo{journal}{Phys. Rev.
  Lett.} \bibinfo{volume}{98} (\bibinfo{year}{2007})
\bibinfo{pages}{262002} [hep-ph/0703120].

\bibitem[{Dittmaier et~al.(2009)Dittmaier, Uwer, and
  Weinzierl}]{Dittmaier:2008uj}
\bibinfo{author}{S.~Dittmaier}, \bibinfo{author}{P.~Uwer},
  \bibinfo{author}{S.~Weinzierl}, \bibinfo{title}{{Hadronic top-quark pair
  production in association with a hard jet at next-to-leading order QCD:
  Phenomenological studies for the Tevatron and the LHC}},
  \bibinfo{journal}{Eur. Phys. J.} \bibinfo{volume}{C59} (\bibinfo{year}{2009})
  \bibinfo{pages}{625--646} [arXiv:0810.0452].

\bibitem[{Melnikov and Schulze(2010)}]{Melnikov:2010iu}
\bibinfo{author}{K.~Melnikov}, \bibinfo{author}{M.~Schulze},
  \bibinfo{title}{{NLO QCD corrections to top quark pair production in
  association with one hard jet at hadron colliders}}, \bibinfo{journal}{Nucl.
  Phys.} \bibinfo{volume}{B840} (\bibinfo{year}{2010})
  \bibinfo{pages}{129--159} [arXiv:1004.3284].

\bibitem[{Sjostrand et~al.(2006)Sjostrand, Mrenna, and
  Skands}]{Sjostrand:2006za}
\bibinfo{author}{T.~Sjostrand}, \bibinfo{author}{S.~Mrenna},
  \bibinfo{author}{P.~Z. Skands}, \bibinfo{title}{{PYTHIA 6.4 Physics and
  Manual}}, \bibinfo{journal}{JHEP} \bibinfo{volume}{05} (\bibinfo{year}{2006})
  \bibinfo{pages}{026} [hep-ph/0603175].

\bibitem[{Corcella et~al.(2002)}]{Corcella:2002jc}
\bibinfo{author}{G.~Corcella}, et~al., \bibinfo{title}{{HERWIG 6.5 release
  note}}  [hep-ph/0210213].

\bibitem[{M. Schulze (2011)}]{Schulze:private}
\bibinfo{author}{M.~Schulze}, \bibinfo{title}{{private communication}}.

\bibitem[{Catani et~al.(1993)Catani, Dokshitzer, Seymour, and
  Webber}]{Catani:1993hr}
\bibinfo{author}{S.~Catani}, \bibinfo{author}{Y.~L. Dokshitzer},
  \bibinfo{author}{M.~H. Seymour}, \bibinfo{author}{B.~R. Webber},
  \bibinfo{title}{{Longitudinally invariant $K_t$ clustering algorithms for
  hadron hadron collisions}}, \bibinfo{journal}{Nucl. Phys.}
  \bibinfo{volume}{B406} (\bibinfo{year}{1993}) \bibinfo{pages}{187--224}.

\bibitem[{Cacciari and Salam(2006)}]{Cacciari:2005hq}
\bibinfo{author}{M.~Cacciari}, \bibinfo{author}{G.~P. Salam},
  \bibinfo{title}{{Dispelling the $N^{3}$ myth for the $k_t$ jet-finder}},
  \bibinfo{journal}{Phys. Lett.} \bibinfo{volume}{B641} (\bibinfo{year}{2006})
  \bibinfo{pages}{57--61} [hep-ph/0512210].

\end{thebibliography}

\end{document}